\documentclass[a4paper, amsfonts, amssymb, amsmath, reprint, showkeys, nofootinbib, twoside]{revtex4-1}

\usepackage{graphics}      
\usepackage{graphicx}      
\usepackage{url}           
\usepackage{bm,color}      
\usepackage{amssymb, amsmath, enumerate, theorem, epsfig,subfigure,setspace}
\usepackage{amsfonts}
\usepackage{bm}

\usepackage[table]{xcolor}

\begin{document}
\title{XY Neural Networks}
\author{Nikita Stroev$^1$ and Natalia G. Berloff$^{1,2}$ }
\email[correspondence address: ]{N.G.Berloff@damtp.cam.ac.uk}
\affiliation{$^1$Skolkovo Institute of Science and Technology, Bolshoy Boulevard 30, bld.1,
		Moscow, 121205, Russian Federation}
\affiliation{$^2$Department of Applied Mathematics and Theoretical Physics, University of Cambridge, Cambridge CB3 0WA, United Kingdom}

\date{\today}%

\begin{abstract}

The classical XY model is a lattice model of statistical mechanics notable for its universality in the rich hierarchy of the optical, laser and condensed matter systems. We show how to build complex structures for machine learning based on the XY model's nonlinear blocks. The final target is to reproduce the deep learning architectures, which can perform complicated tasks usually attributed to such architectures:  speech recognition, visual processing, or other complex classification types with high quality. We developed the robust and transparent approach for the construction of such models, which has universal applicability (i.e. does not strongly connect to any particular physical system), allows many possible extensions while at the same time preserving the simplicity of the methodology.

\end{abstract}

\maketitle

\section{Introduction}
\label{Introduction}
The growth of modern computers suffers from many limitations, among which is Moore's law \cite{moore1965cramming,moore2000readings}, an empirical observation about the physical volume-induced limitation on the number of transistors in integrated circuits, increased energy consumption connected with the growth of the performance, intrinsic issues arising from the architecture design like the so-called von Neumann bottleneck \cite{von1993first}, etc. The latter refers to the internal data transfer process built according to the von Neumann architecture and usually denotes the idea of computer system throughput limitation due to the characteristic of bandwidth for data coming in and out of the processor. Von Neumann bottleneck problem was addressed in various ways \cite{edwards2016eager,naylor2007reduceron} including managing multiple processes in parallel, different memory bus design, or even considering conceptual "non-von Neumann" systems \cite{lent2016molecular,shin2019heterogeneous}.

The inspirations for such information-processing architectures comes from different sources, artificial or naturally occurring. This concept is usually referred to as neuromorphic computation \cite{schuman2017survey,furber2016large}, which competes with state-of-the-art performance in many specialized tasks like speech and image recognition, machine translation and others. Additionally, it has the property of distributed memory input and computation compared to the linear, and conventional computational paradigm \cite{sompolinsky1988statistical}. The artificial neural networks (NNs) demonstrated many advantages compared to classical computing \cite{lecun2015deep}. NN approach offers a solution to the von Neumann bottleneck but gives an alternative view of solving a particular task and the entire idea of computation.  The associative memory model is the critical paradigm in the NN theory and is intrinsically connected with NN architectures.

An alternative approach to the particular computational task is to encode it into the NN coefficients and solve the specific cost function minimization instead of performing the predefined sequence of computations (e.g. logical gates). The nodes' final configuration through a nonlinear evolution of the system corresponds to the initial problem's solution. One of the familiar and famous models of the NNs is  Hopfield networks \cite{hopfield1986computing} which serve as content-addressable or associative memory systems with binary threshold nodes. Its mathematical description is similar to the random, frustrated Ising spin glass \cite{stein2013spin}, for which finding the ground state is known to be an NP-hard problem \cite{barahona1982computational}. It reveals an interplay between the condensed matter and computer science fields while offering various physical realizations in optics, lasers and beyond.

The Ising Hamiltonian can be treated as a degenerate XY Hamiltonian. The classical XY model, sometimes also called classical rotator or simply O(2) model, is a subject of extensive theoretical and numerical research and appears to be one of the basic blocks in the rich hierarchy of the condensed matter systems \cite{chaikin1995principles,kosterlitz1974critical,gupta1988phase,gawiec1991numerical,kosterlitz1999numerical}. It has an essential property of universality, which allows one to describe a set of mathematical models that share a single scale-invariant limit under renormalization group flow with the consequent similar behaviour \cite{svistunov2015superfluid}. Thus, the model's properties are not influenced by whether the system is quantum or classical, continuous or on a lattice, and its microscopic details. These types of models are usually used to describe the properties of magnetic materials or the superfluid state of matter \cite{svistunov2015superfluid}. Moreover, they can be artificially reproduced through intrinsically different systems with the latest advances in experimental manipulation techniques \cite{buluta2009quantum,aspuru2012photonic,kjaergaard2020superconducting,hauke2012can}. 

Reproducing the Ising, XY, or NN architecture model using a particular setup allows one to utilize a particular physical system's benefits. Extending this problem to the physical domain on the hardware level, we can additionally solve other computing problems by decreasing the energy consumption or volume inefficiency or even increasing the processing speed. The  NN implementation has been previously discussed in the optical settings \cite{montemezzani1994self,psaltis1995holography}, all-optical realisation of reservoir computing \cite{duport2012all} or using silicon photonic integrated circuits \cite{shen2017deep} and other coupled classical oscillator systems \cite{csaba2020coupled}. Exciton-polariton condensates are one of such systems (together with many others, for instance, plasmon-polaritons \cite{frank2012coherent,frank2013non}) that led to a variety of applications such as a two-fluid polariton switch \cite{de2012control},
all-optical exciton-polariton router \cite{marsault2015realization},
polariton condensate transistor switch \cite{Gao_transistor} and
polariton transistor \cite{Ballarini_transistor}, which can also be realized  at  room-temperatures \cite{zasedatelev2019room}, spin-switches \cite{Amo_switch}
and the XY-Ising simulators \cite{berloff2017realizing,lagoudakis2017polariton, kalinin2018simulating}. The coupling between different condensates can be engineered with high precision while the phases of the condensate wavefunctions arrange themselves to minimize the XY Hamiltonian. 

There has been recent progress in developing the direct hardware for the artificial NNs, emphasising optical and polaritonic devices.  Among them,  there is  the concept of reservoir computing in the Ginzburg-Landau lattices, applicable to the description of a broad class of systems, including exciton-polariton lattices \cite{opala2019neuromorphic}, and the consequent experimental realization with the increases in the recognition efficiency and high signal processing rates \cite{ballarini2020polaritonic}. The implementation of the backpropagation mechanism through neurons in an all-optical training manner in the field of optical NNs through the approximation mechanism with the pump-probe passive elements resulted in the equivalent state-of-the-art performance \cite{guo2019end}. The near-term experimental platform for realizing an associative memory based on spinful bosons coupled to a degenerate multimode optical cavity enjoyed advanced performance due to the system tuning. It led to specific dynamics of neurons in comparison with the conventional one \cite{marsh2020enhancing}.

This paper shows how to perform the NN computation by establishing the correspondence between the XY model's nonlinear clusters that minimize the corresponding XY Hamiltonian and the basic mathematical operations, therefore,  reproducing neural network transformations.  Moreover, we solve an additional set of problems using the properties offered explicitly by the exciton-polariton systems. The ultra-fast characteristic timescale for condensation (of the order of picoseconds) allows one to decrease the computational time significantly. The intrinsic parallelism, based on working with  multiple operational nodes, can enhance the previous benefits compared to the FPGA or tensor processing units. The nonlinear behaviour can be used to perform the approximate nonlinear operations. This task is particularly computationally inefficient on the conventional computing platforms where a local approximation in evaluation is commonly used.  
Since there are many trivial ways of utilizing the correspondence between the Ising model and the Hopfield NN in terms of the discrete analogue variables, instead, we use the continuous degrees of freedom of the XY model. By moving to continuous variables, we address the volume efficiency problem since converting the deep NNs to the quadratic binary logic architecture meets significant overhead in the number of variables while increasing the range of the coupling coefficients. The former may be out-of-reach for the actual physical platforms, while the latter increases the computational errors.

With the motivation to transfer the deep learning (DL) architecture into the optical or condensed matter platform, we show how to build complex structures based on the XY model's nonlinear blocks that naturally behave in a nonlinear way in the process of reaching the equilibrium. The corresponding spin Hamiltonian is quite general because it can be engineered with many condensed matter systems acting as analogue simulators \cite{pargellis1994planar,struck2013engineering,berloff2017realizing}.

The key points of our paper are:
\begin{itemize}
    \item We show how to realize the basic numerical operations using some small-size XY networks, which  minimize the XY Hamiltonian clusters' energy.
    \item We obtain the complete set of operations  sufficient to realize different combinations of the mathematical operations and to map the deep NN architectures into XY models.
    \item We demonstrate that this approach can be applied to the existing special-purpose hardware capable of reproducing the XY models.
    \item We present the general methodology based on the simple nonlinear systems, which can be extended to other spin Hamiltonians, i.e. with different interactions or models with additional degrees of freedom and involving quantum effects.
\end{itemize}

This paper is organized as follows. Section \ref{Basic XY equilibrium blocks} is devoted to describing our models' basic blocks and introducing notations used. Section \ref{Demonstration of the function approximations} contains the demonstration of our approach's effectiveness in approximating simple functions. Section \ref{Neural Networks benchmarks} extends this approach to the small NN architectures. Section \ref{Transfering Deep Learning architecture} considers the extensions to the deep architectures with a particular emphasis on various nonlinear functions implementations. Subsection \ref{Exciton-polariton setting} is dedicated to the particular exciton-polariton condensed matter system as a potential platform for implementing our approach. Conclusions and future directions are given in the final Section \ref{Conclusions and future directions}.
\textit{Supplementary information} provides some technical details about the NN parameters, approximations, and corresponding optimization tasks.

\section{Basic XY equilibrium blocks}
\label{Basic XY equilibrium blocks}

This section is devoted to describing the basic blocks of the XY NN with a particular emphasis on the DL architecture. DL is usually defined as a part of the ML methods based on the artificial NN with representation learning and proven to be effective in many scientific domains \cite{goodfellow2016deep,lecun2015deep,szegedy2015going,he2016deep}. It appears to be applicable in many areas ranging from applied fields such as chemistry and material science to fundamental ones like particle physics and cosmology \cite{carleo2019machine}.

The DL is typically referred to as a black box \cite{zdeborova2020understanding,deng2014deep}. When it comes to DL's application, one usually asks questions about adapting the DL architectures to the new problems, how to interpret the results and how to quantify the outcome errors reliably. Leaving these open problems behind, we formulate a more applied task. To build DL architectures, we want to transfer the pretrained parameters into a realization of a nonlinear computation. One of the mathematical core ideas in machine learning (ML) architectures is the ability to build hyperplanes on each neuron output, which taken together allows one to approximate the input data efficiently and adjust it to the output in the case of supervised learning (for example, in the classification tasks \cite{kotsiantis2007supervised}). This procedure can be paraphrased as the feature engineering that before the modern DL approaches and the available computational resources was performed in a manual way \cite{turner1999conceptual}.

Building hardware that performs the hyperplane transformation with a specific type of a nonlinear activation function with a given precision allows one to separate input data points and present building block operations for more complex tasks. Studying the hierarchical structures with such blocks leads to constructing more complex architectures capable of performing more sophisticated tasks.

Decomposing the nonlinear expressions common in the ML such as $\tanh(w_0 x_0 + w_1 x_1 + … + w_n x_n + b),$ produces a set of mathematical operations, which we need to approximate with our system. These are nonlinear operation $\tanh$, which is conventionally called an activation function, the multiplication of the input variables $x_i$ by the constant (after training) coefficients $w_i$ (also called weights of a NN) and the summation operation (with the additional constant $b$, called bias).

The activation function is an essential aspect of the deep NNs, bringing nonlinearity into the learning process. The nonlinearity allows modern NNs to create complex mappings between the inputs and outputs, vital for learning and approximating complex data with high dimensionality.
Moreover, the nonlinear activation functions afford backpropagation due to the smooth derivative of those functions. They normalize each neuron's output, allowing one to stack multiple layers of neurons to create a deep NN. The functional form of the nonlinear function is zero centred with the saturation effect, which leads to a certain response for the inputs that take strongly negative, neutral, or strongly positive values, not to mention the information-entropic foundation of its derivative form and other interesting properties \cite{taylor1993information,fyfe2000artificial}. We will see that approximating this operation is straightforward with the XY networks.

Firstly, we introduce the list of simple blocks corresponding to the set of operations necessary to realize the nonlinear activation function, which can be obtained by manipulating the small clusters of spins with underlying U(1) symmetry. These clusters minimize the XY Hamiltonian: 
\begin{equation}	
H= \sum_{i=1}^N\sum^{N}_{j=1} J_{ij} \cos(\theta_i-\theta_j),
\label{xy_hamiltonian}		
\end{equation}
where $i$ and $j$ goes over $N$ elements in the system, $J_{ij}$ is the interaction strength between $i^{th}$ and $j^{th}$ spins represented by the classical phases $\theta_i \in [-\pi,\pi]$.  If we take several spins as inputs ${\theta_i}$ in such a system and consider the others as outputs ${\theta_k}$, then we can treat the whole system as a nonlinear function which returns $\underset{\theta_k}{\arg\min}  H(\{\theta_i\},\{\theta_k\})$ values due to the system equilibration into the steady-state. In some cases the ground state is unique, other cases can produce multiple equilibrium states.

\begin{figure}
\centering
\includegraphics[width=0.4\textwidth]{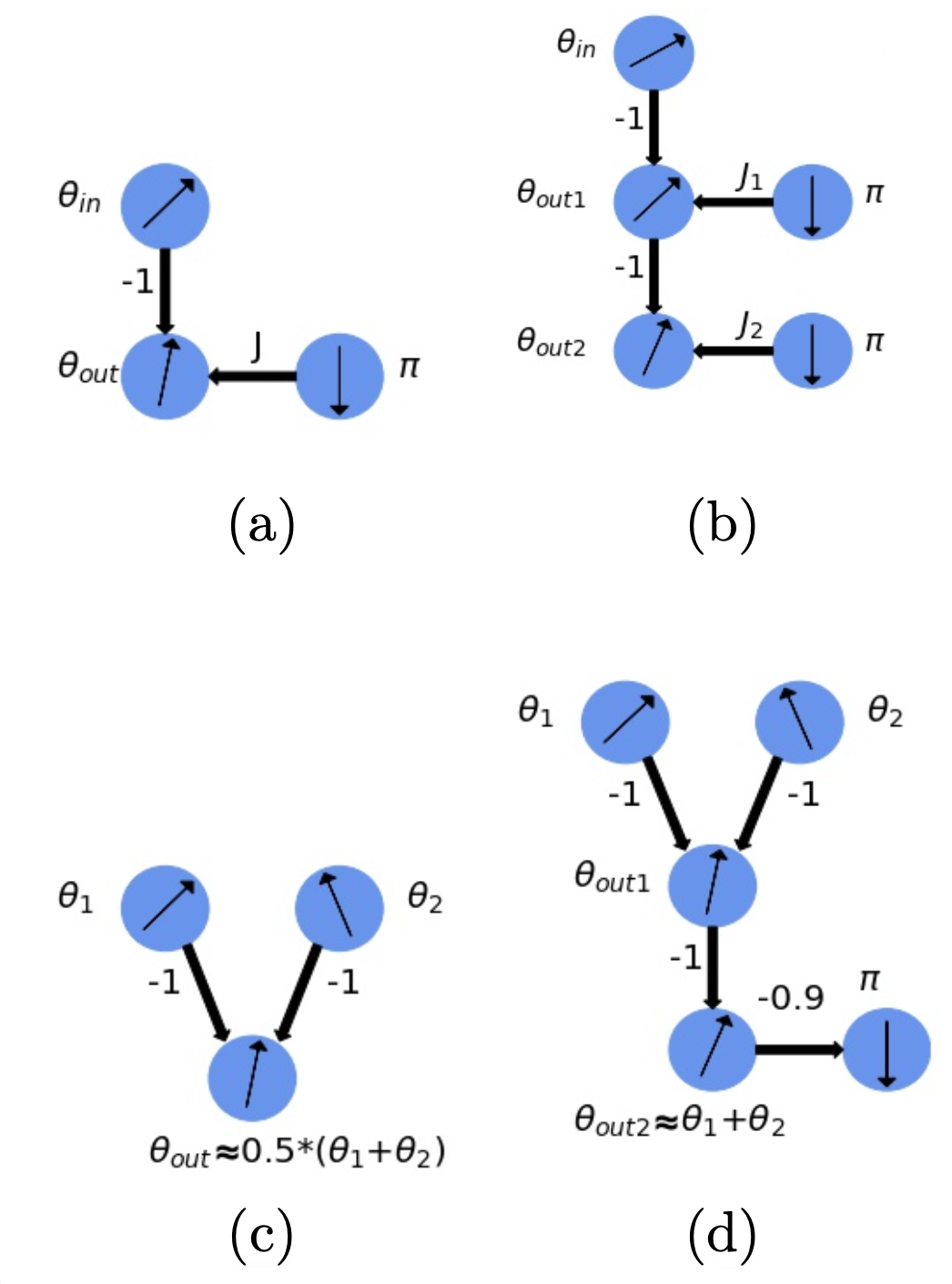}
\caption{Several examples of basic blocks and their combinations used in the XY NN architectures. 
(a) The block performing the function $F((\theta_{in}|-1),(\pi|J))$ with one input spin and one reference/control spin with imposed $\pi$ value, all coupled with the output by the ferromagnetic $-1$ and $J$ ($J=1$  is used on the picture).
Depending on $J$ we can realise the operations that approximate the multiplication by the constant $k$ so that $\theta_{out} \approx 1.5\tanh{4\theta_{in}} + 0.5\theta_{in}$ with $J=-0.9$.
(b) Two blocks representing $F(F((\theta_1|-1),(\pi|J_1))|-1),(\pi|J_2))$. 
(c) The block $F((\theta_{1}|-1),(\theta_{2}|-1))$ for the the half sum of two variables $\theta_1$ and $\theta_2$.
(d) Two blocks performing the function
$F(F((\theta_1|-1),(\theta_2|-1))|-1),(\pi|-0.9))$.
Some of the response functions for these blocks are presented in Fig.\ref{basic_plots}.}
\label{basic_blocks}
\end{figure}

\begin{figure}
\centering
\includegraphics[width=0.48\textwidth]{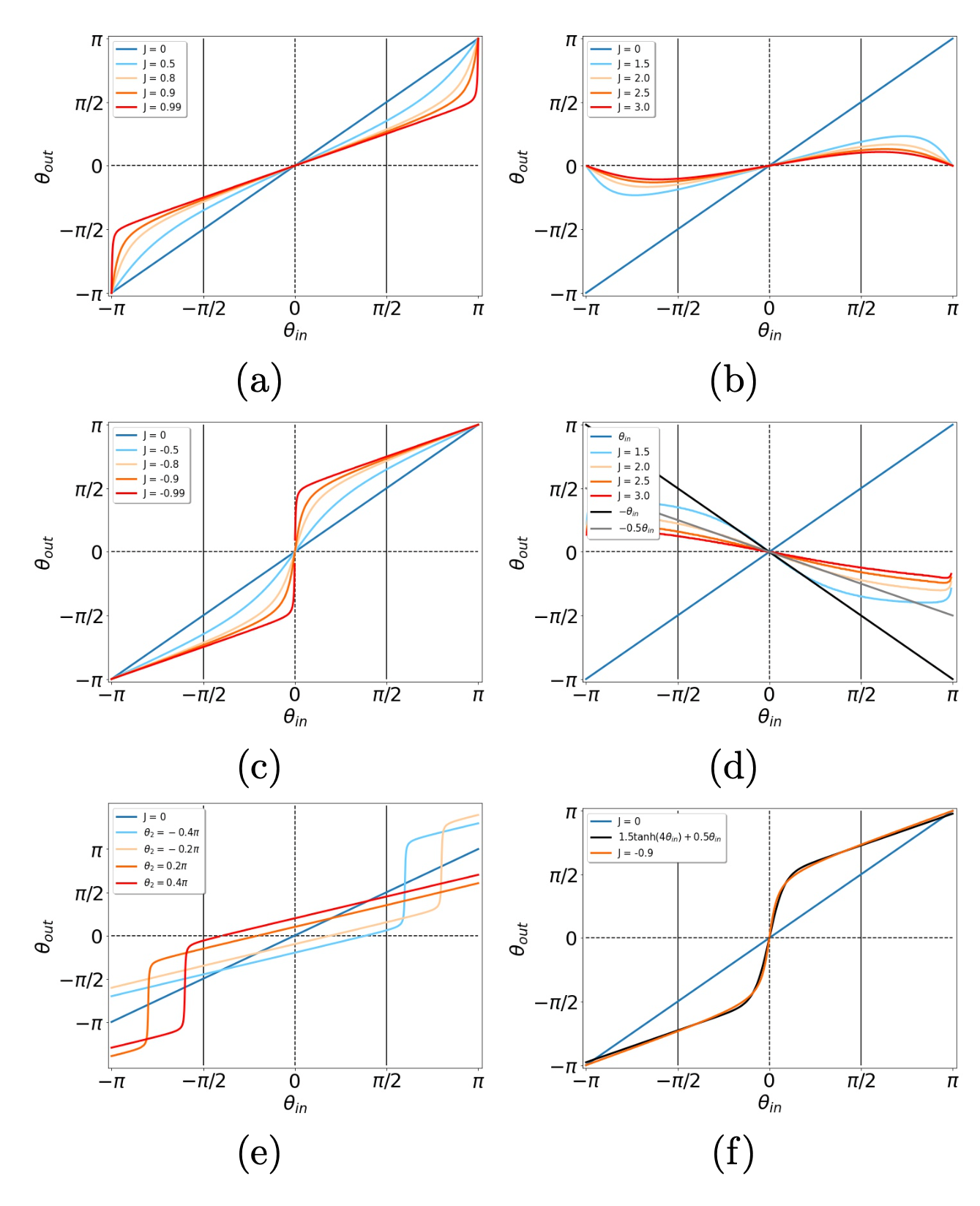}
\caption{Several examples of input-output relations for the basic blocks and their combinations used in the XY NN architectures.
(a) The parametrised family of $F((\theta_{in}|-1),(\pi|J))$ functions, that corresponds to the basic blocks. Depending on $J$ coupling strength parameter we can realise the multiplication by arbitrary $k$.
(b) The graphs of $F((\theta_{in}|-1),(\pi|J))$ functions for various values of $J$ illustrating  the  the multiplication by  small values of $k$.
(c) The graphs of $F((\theta_{in}|-1),(\pi|J))$ functions for smaller values of  $J<0$. 
(d) The graphs of $F^{3}_3(F_2(F^{2}_1(\theta_{in})|J))$ functions, where $F_1(\theta_{in}) = F((\theta_{in}|-1),(\pi|0.9)), F_2(\theta_{in}|J) = F((\theta_{in}|1),(\pi|J)), F_3(\theta_{in}) = F((\theta_{in}|-1),(\pi|-0.2))$, showing different negative outputs.
(e) The graphs of $F((\theta_{1}|-1),(\theta_{2}|-1))$, implementing the block shown on  Fig.~\ref{basic_blocks}(c), which approximates half sum of input variables.
(d) The graphs of $1.5\tanh(4\theta_{in})+0.5\theta_{in}$ and  $F((\theta_{in}|-1),(\pi|-0.9))$.
}
\label{basic_plots}
\end{figure}

It is useful to consider the analytical solution to such kind of task, describing the function with one output and several input variables. We consider the system with $N$ spins: $\theta_i, i=1,..,N-1$ are input spins and $\theta_N$ is the output spin coupled with the input spins by the strength coefficients $J_{i} \equiv J_{iN}$. The system Hamiltonian can be written as $H= \sum^{N-1}_{i} J_{i} \cos(\theta_i-\theta_N)$. By expanding $H$ as $\sum^{N-1}_{i} J_{i} \cos(\theta_i-\theta_N) = \sum^{N-1}_{i} J_{i} \cos \theta_i \cos\theta_N + \sum^{N-1}_{i} J_{i} \sin\theta_i \sin\theta_N$ we can solve for the minimizer $\theta_{N}$:
\begin{equation}	
\theta_N \equiv  F((\theta_1|J_1),(\theta_2|J_2),...,(\theta_{N-1}|J_{N-1})) =\nonumber
\end{equation}
\begin{equation}
=-\operatorname{sign}B \left(\frac{\pi}{2} + \arcsin\frac{A}{\sqrt{A^2 + B^2}}\right),
\label{xy_hamiltonian_analytics}		
\end{equation}
 where $A=\sum^{N-1}_{i} J_{i} \cos \theta_i$ and $B=\sum^{N-1}_{i} J_{i} \sin \theta_i$. Alternatively, this formula can be rewritten through the complex analog of the order parameter $C = \sum^{N-1}_{i} J_{i} e^{i \theta_{i}} = A + iB$ in the following way: $\theta_N = F(\{\theta_i|J_i\}) = -\operatorname{sign}(\operatorname{Im} C) \left(\pi - \operatorname{Arg} C \right)$. We present several basic blocks in Fig.~\ref{basic_blocks} and the outcomes of the functions responses in Fig.~\ref{basic_plots}.

We will use the notation introduced in Eq.~(\ref{xy_hamiltonian_analytics}) to describe both the activation function and the graph cluster of spins below. 
We will use the recurrent notation where the output of the first block serves as the input to the next one, for example $F(F((\theta_1|J_1),(\theta_2|J_2))|J_3),(\theta_4|J_4))$.

To describe the iterative implementation of many ($k$) identical blocks, where the input is defined in terms of the output of the previous same block, we rewrite the recurrent formula for one argument as: $(F_1 \circ F_1 \circ ... \circ F_1) (\theta_{in}) \equiv  F_1(F_1(...F_1(\theta_{in}))) \equiv  F^{k}_1(\theta_{in})$, where $F_1(\theta_{in}) = F((\theta_{in}|J_1),(\theta_2|J_2))$ is a certain block with the predefined parameters.
We separate all possible blocks of spins into several groups and consider them below in more detail.

The phase $\theta_i$ are in $[-\pi,\pi]$, however, for an efficient approximation of the operations (summation, multiplication and nonlinearity) we need to limit the domain to   $[-\pi/2,\pi/2]$ which we refer to as the \textit{working domain}. Additionally, we need to guarantee that the values of the \textit{working spins} (which are not fixed and are influenced by the system input, thus serving as analogue variables) are located within the limits of the working domain. This will be implemented below. 
Next, we consider the implementation of the elementary operations.

\textit{- Multiplication by the constant value $k>0$}. Connecting the input spin with the output spin by the “ferromagnetic” coupling $J=-1$ will lead to the input spin's replication. In this way, we can transmit the spin value from one block to another. Changing the value of the output spin can be achieved in many ways. The addition of another spin with a different value and coupling it to the output spin with a constant coupling $J$ is one such possibility (for example, with imposed $\pi$ value, which we will refer to as a \textit{reference/control spin}). If $J$ is in $[0,1)$ (relative to $-1$ coupling between $\theta_{in}$ and $\theta_{out}$), then the reference spin influences the output spin value with the effective “repulsion” and thus, depending on the relative coupling strength, decreases the output spin value (see Fig.~\ref{basic_plots}(a,b) and the corresponding cluster configuration on Fig.~\ref{basic_blocks}(a)). The resulting relation between the input and output spin values can be a good approximation to the multiplication by certain values lying in the $[0,1]$ range. The block corresponding to the implementation of $F((\theta_{in}|-1),(\pi|1))$
has a peculiarity in case of $\theta_{in} = 0$, which allows the output to take any value due to the degeneracy of the ground state. To overcome this degeneracy, we choose $J=0.99$. 

For $k>1$ we can use a ferromagnetic coupling $J<0$ (see Fig.~\ref{basic_plots}(c) and the corresponding cluster configuration on Fig.~\ref{basic_blocks}(a)). However, the positive values of $J$ are more reliable for the implementation since the output functions have small approximation errors (see \textit{Supplementary information} for the exact values of this error and further clarification). We can replace the multiplication by a large factor by  the multiplications by several  smaller factors to reduce the final accumulated error. We can guarantee the uniqueness of the output since the clusters are small, and the output is defined by Eq.~(\ref{xy_hamiltonian_analytics}), which gives the unique solution.

\textit{- Nonlinear activation function}. 

The function $F((\theta_{in}|-1),(\pi|-0.9))$ is similar to the hyperbolic $\tanh$ function (see Fig.~\ref{basic_plots}(c) and the  \textit{Supplementary information} for the exact difference).
There are two ways of using such a transformation as an activation function.

1) We can use the similarity between the values of the $F((\theta_{in}|-1),(\pi|-0.9))$ and the function $1.5\tanh(4\theta_{in})+0.5\theta_{in}$ (see Fig.~\ref{basic_plots}(f)).
In general, many nonlinear functions can be used in NNs. Usually, a minor modification in the functional form of the nonlinear activation function does not change the network's overall functionality, with the additional training procedures of NNs in some architectures. We can train the NN initially with the $1.5\tanh(4\theta_{in})+0.5\theta_{in}$ function so that in the final transfer, it will not be necessary to adjust the spin system to approximate the given function.

2) We can use the  similarity with the approximate hyperbolic tangent function within the XY spin cluster. In other words, to execute $\tanh(\theta_{in})$, we have to perform $(F((0.25 \theta_{in}|-1),(\pi|-0.9)) - 0.5\theta_{in})/1.5$ function using the spin block operations. This option will be used below.

\textit{- Multiplication by the constant value $k=-1$}.
The main difficulty of this  operation is in finding the set of parameters for the spin block where  $\frac{\partial F}{\partial \theta_{in}} < 0$ holds.  $F((\theta_{in}|1),(\pi|J))$  is a good example of such a block.
To perform the multiplication by $k=-1$, we need to embed the whole working domain into the region where the presented inequality is valid and return these values with the multiplication by $k>0$ factor.
One final realization can be represented as
$F^{3}_3(F_2(F^{2}_1(\theta_{in})))$ function, where $F_1(\theta_{in}) = F((\theta_{in}|-1),(\pi|0.9)), F_2(\theta_{in}) = F((\theta_{in}|1),(\pi|J)), F_3(\theta_{in}) = F((\theta_{in}|-1),(\pi|-0.2))$.

\textit{- Summation}. The function $F((\theta_{1}|-1),(\theta_{2}|-1))$ gives a good approximation to the half sum $(\theta_{1}+\theta_{2})/2$. This block is presented in Fig.~\ref{basic_blocks}(c) and the cross-sections of the surface defined by the function of two variables $F((\theta_{1}|-1),(\theta_{2}|-1))$ are plotted in Fig.~\ref{basic_plots}(e). The plots show that the spin system realizes the half sum of two spin values with a minimum discrepancy compared to the target function on a working domain. One can multiply the final result by two using previously described multiplication to achieve an ordinary summation. In general, such a type of summation can be extended on multiple spins $N>2$, in a similar way of connecting them to the output spin, with the final value of $(\theta_{1}+..+\theta_{N})/N$.

Summarizing, we presented a method of approximating the set of mathematical operations, necessary for performing the $\tanh(w_0 x_0 + w_1 x_1 + … + w_n x_n + b)$ function, using the XY blocks  described by Eq.~$\ref{xy_hamiltonian_analytics}$.  The output spin value of each block is formed when a global equilibrium is reached in the physical system with the speed that depends on particular system and its parameters. We kept the universality of the model, which allows one to implement this approach using various XY systems.
We did not use any assumptions about the nature of the classical spins, their couplings, or the manipulation techniques, however, the forward propagation of information requires directional couplings.
As the blocks corresponding to elementary operations are added one after another, the new output spins and new reference spins should not change the values of  the output spins from the previous block. The directional couplings that affect the output spins of the next block but not the output spins of the previous block satisfy this requirement. Many systems can achieve directional couplings. For instance, in optical systems the couplings are constructed by redirecting the light with either free- space optics or optical fibres to a spatial light modulator (SLM). At the SLM, the signal from each node is multiplexed and redirected to other nodes with the desired directional coupling strengths \cite{kalinin2020polaritonic}.

\section{One-dimensional function approximations}
\label{Demonstration of the function approximations}

This section illustrates the efficiency of the proposed approximation method on  one-dimensional functions of intermediate complexity by considering two examples of mathematical functions and their decomposition into the basis of nonlinear operations.

For illustration, we choose two  nontrivial functions (one is monotonic and another is non-monotonic). The use of the methodology for more complex functions in higher dimensions is straightforward. In the next section, we will show the method efficiency for two-dimensional data-scientific toy problems.

We consider two functions
\begin{equation}	
{\cal F}_1=0.125F_t(1.2x)+0.125F_t(0.5(x+1.4)),
\label{app_function1}		
\end{equation}
and 
\begin{equation}	
{\cal F}_2=0.125F_t(0.5(x-1.2))-0.03125F_t(0.5(x+1.2)),	
\label{app_function2}		
\end{equation}
where $F_{t}(x) = 1.5\tanh(4x)$. Note, that arbitrary functions can be obtained using a linear superposition of scaled and translated basic functions $F_{t}(x)$.

\begin{figure} 
\centering
\includegraphics[width=0.5\textwidth]{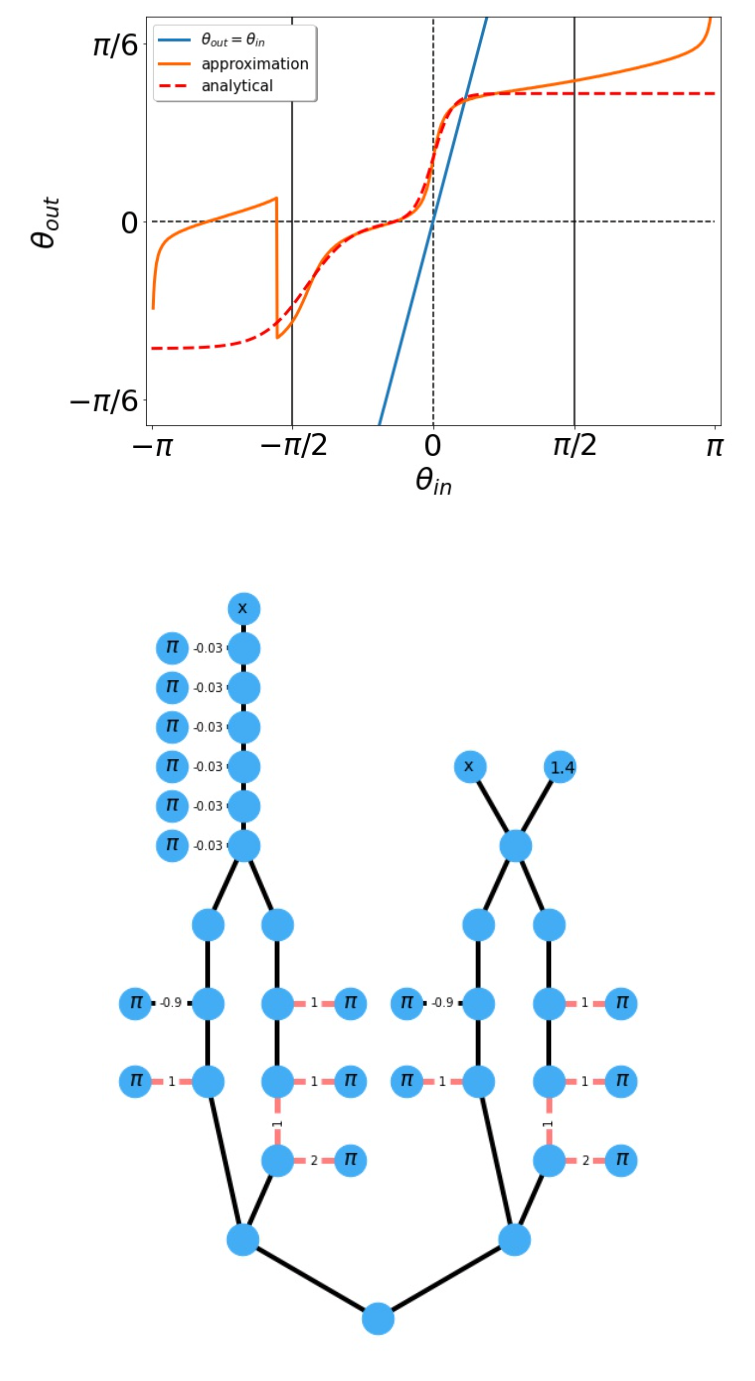}
\caption{Top: The demonstration of the approximation quality, obtained by using the nonlinear XY spin clusters. The analytical monotonic function (red dashed line) is given by Eq.(\ref{app_function1}), the orange line is the approximation, the blue line is $\theta_{out}=\theta_{in}$.
Bottom: The graph structure, representing the basic mathematical operation in the Eq.(\ref{app_function1}), given by the blocks, discussed in Section \ref{Basic XY equilibrium blocks}. The input variables are mapped into the top spins, after which the cluster is equilibrated before performing the next operation.  The blue empty nodes are working spins that  change according to the variables at higher block. The blue nodes with the fixed $\pi$-value are reference/control spins. The black edges without the notation have the fixed relative strength $-1$, while for others the coupling is written 
on the edges explicitely. The red colour of the edge represents the positive relative coupling strength $J$. The bottom spin gives the value of the coded function.}
\label{approximation1}
\end{figure}

\begin{figure} 
\centering
\includegraphics[width=0.5\textwidth]{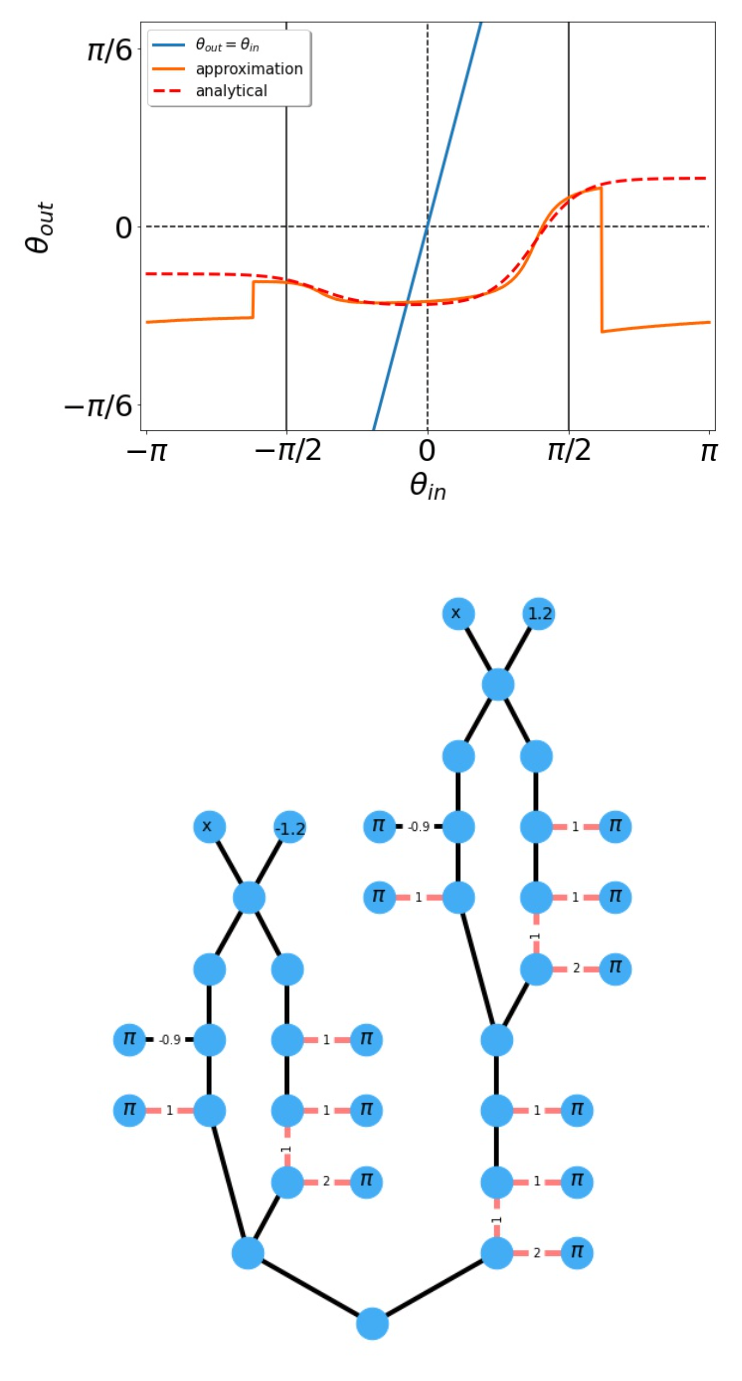}
\caption{Top: The demonstration of the approximation quality, obtained by using the nonlinear XY spin clusters. The analytical non-monotonic function (red dashed line) is given by  Eq.(\ref{app_function2}), the orange line is the approximation, the blue line is the linear identity relation. Bottom: The graph structure, representing the essential mathematical operation in  Eq.(\ref{app_function2}), given by the blocks, discussed in Section \ref{Basic XY equilibrium blocks}. The notation used for describing the graph parameters is the same as in Fig.~\ref{approximation1}.}
\label{approximation2}
\end{figure}

The comparison of the XY blocks approximations and the target functions are given in Fig.~\ref{approximation1} and Fig.~\ref{approximation2} demonstrating a a good agreement  in the working domain. We also plot the explicit structures of  the corresponding XY spin clusters showing  a small overhead on the number of spins used per operation.

\section{Neural Networks benchmarks}
\label{Neural Networks benchmarks}

In this section, we test the XY NN architectures and check their effectiveness using typical benchmarks. For simple architectures, the classification of predefined data points perfectly suits this goal. We consider standard two-dimensional datasets, which conventionally referred to as ‘moons’ and ‘circles’ and can be generated with \textit{Scikit-learn} tools \cite{pedregosa2011scikit}. An additional useful property of such tasks is that they are easy for manual feature engineering.

\begin{figure} 
\centering
\includegraphics[width=0.45\textwidth]{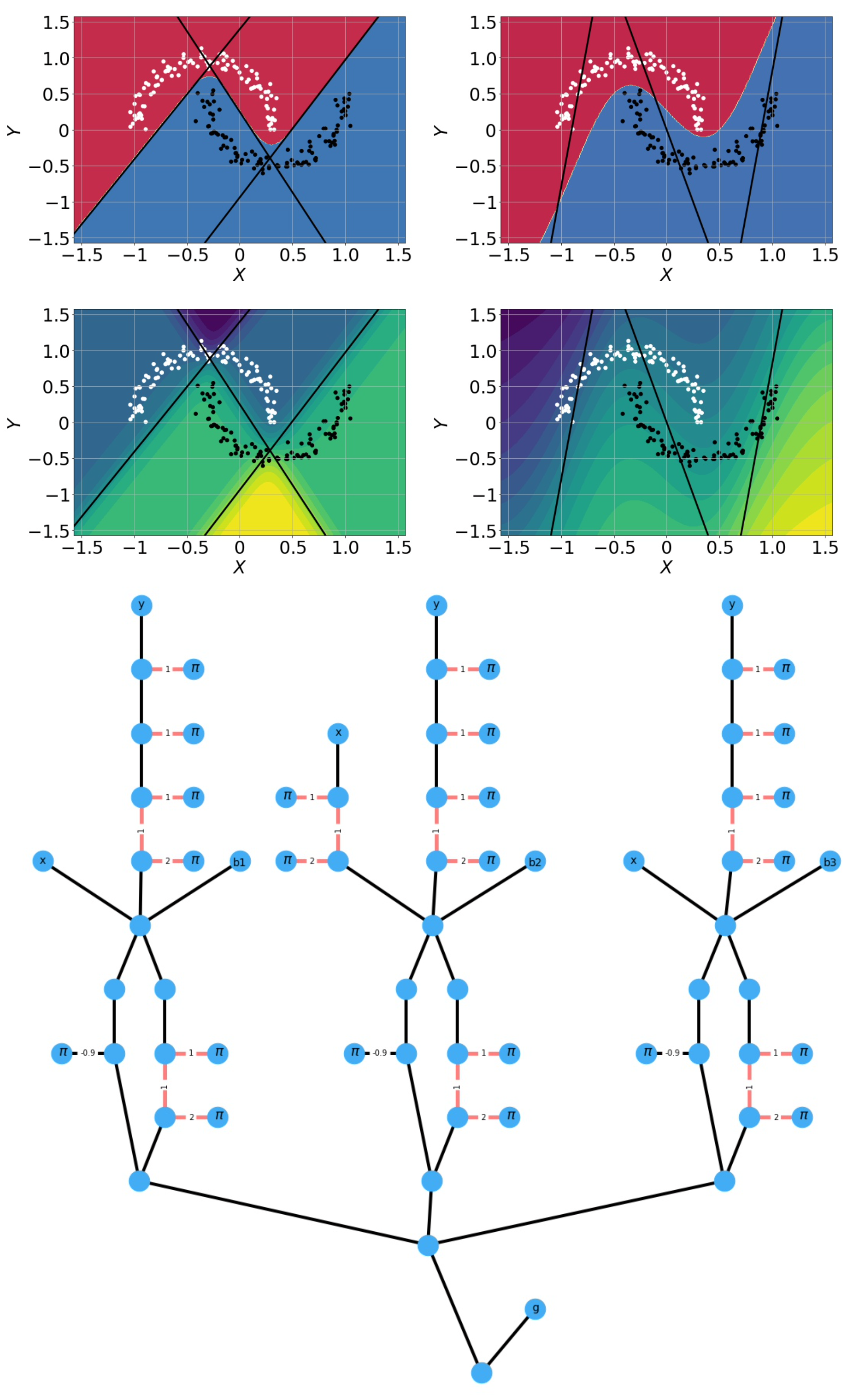}
\caption{Decision boundaries of the simple (2,3,2) NN on the left and approximated ones for the XY NN on the right on toy 2-D moons data set. Black lines represent the bounds for automatically found features in the middle layer of classical NN, the approximated features adjusted for this specific task are located in the right picture.
Middle row: The isosurface for the one particular chosen feature for typical NN and its corresponding matched XY NN last variable isosurface.
The parameters of both NN and XY NN architectures can be found in the \textit{Supplementary Information}. We can state that the XY model can give a good approximation of the classical NN architectures with not ideally smooth representations, but requires small adjustments of their parameters, due to difficulties with representing sharp geometric figures (which in general can be multidimensional).
Bottom: The corresponding graph structure.}
\label{moon_ds}
\end{figure}

First, we train a simple NN. The parameters for the training setup are given in  \textit{Supplementary Information} section. The final performance demonstrates perfect accuracy in both datasets. 
The architecture consists of two neurons' input layer, one hidden layer with three neurons for each feature and $\tanh$ activation function. The output layer consists of two neurons, which are transformed with $\operatorname{Softmax}$ function. The corresponding weights for both cases are given in  \textit{Supplementary Information} section. Fig.~\ref{circle_ds} and Fig.~\ref{moon_ds} show the decision boundaries with the given pretrained architectures and the landscape of one of the final neuron visualizations together with the data points.

Since we are focused on transferring the described architectures into the XY spin cluster system, we consider the basic architecture adjustment on the example of one feature. Suppose we have the expression $\tanh(w_1 x + w_2 y + b)$. To repeat the chosen strategy $2)$ of approximating \textit{the nonlinear activation function}, we rewrite the coefficients $w_1,w_2$ and $b$  as, say, $w_i\rightarrow (N/K)[(K/N)w_i],$ where $K$ is a parameter chosen  to increase the accuracy of each computation.
We approximate the square brackets' operations using the building blocks from the Section \ref{Basic XY equilibrium blocks}. The factor $N=3$ will be cancelled by the value $1/N$ during the summation of $N$ spins, while the factor $K=4$ in the denominator will be taken into account during the operation of the function $F((\theta_{in}|-1),(\pi|-0.9))$, which is close to the $\tanh 4\theta_{in}$. 
The resulting procedure achieves good performance seen in Fig.~\ref{circle_ds}, while the details are  provided further in  \textit{Supplementary information} section.

\begin{figure} 
\centering
\includegraphics[width=0.5\textwidth]{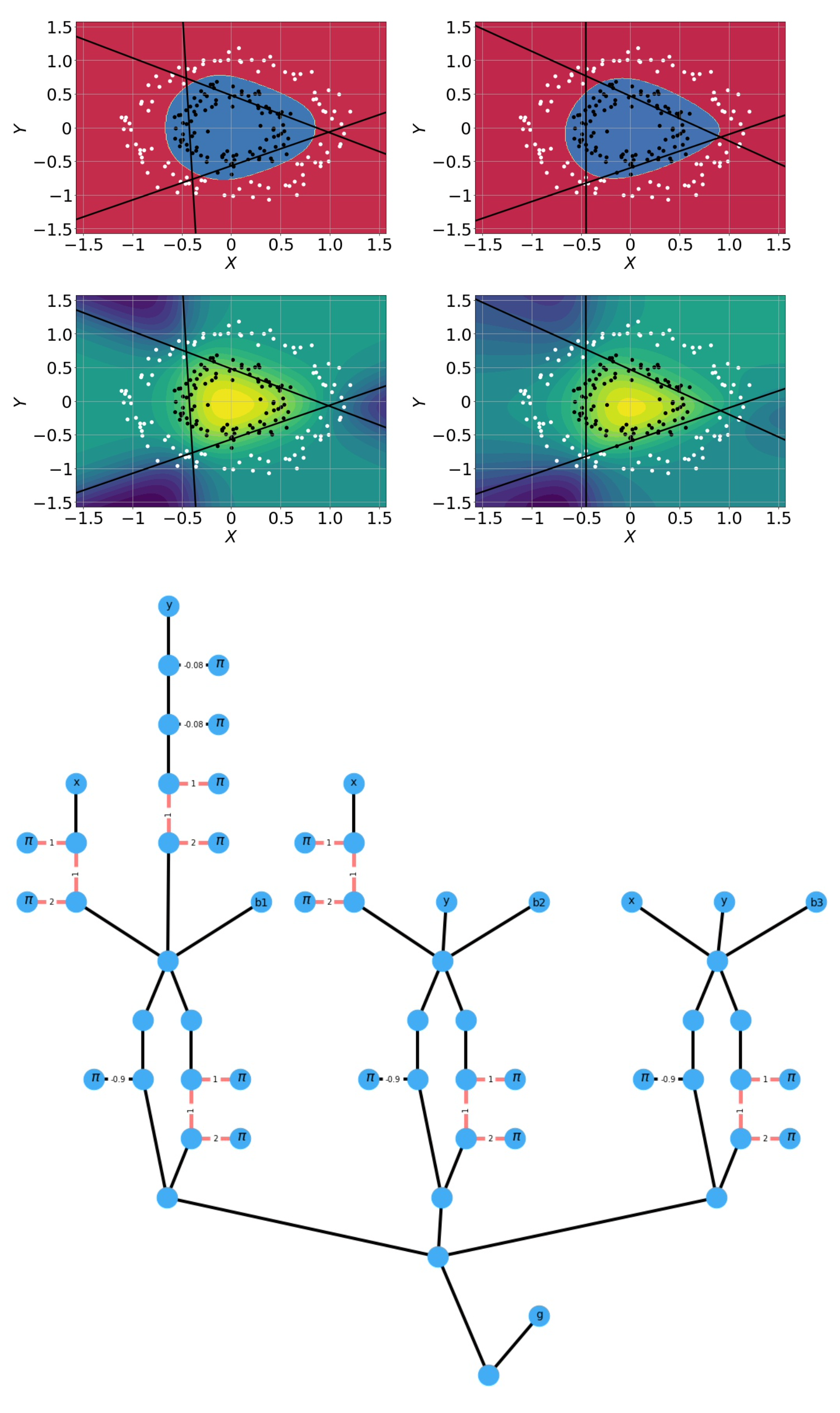}
\caption{Top row: Decision boundaries of the simple (2,3,2) NN on the left and approximated ones for the XY NN on the right on toy 2-D circle data set. Black lines represent the bounds for automatically found features in the middle layer of classical NN, and the approximated features are shown on the right picture.
Middle row: The isosurface for the one particular chosen feature for typical NN and its corresponding matched XY NN last variable isosurface.
The parameters of both NN and XY NN architectures can be found in the \textit{Supplementary Information}.
Bottom: The corresponding graph structure.}
\label{circle_ds}
\end{figure}

The final $\operatorname{Softmax}$ function in the original NN serves as the comparison function of two features to achieve the final decision boundary's smooth landscape. We can omit this function and replace it with a simpler expression $x-y$. To achieve the binary decision boundaries, one can exploit the block performing $F((\theta_{in}|-1),(\pi|-0.9))$ several times to place the final spin value either close to $\pi/2$ or $-\pi/2$. In this way, we adjusted architecture that performs the same functions as the described simple NN on a toy model.  The final decision boundary of the XY NN approximation can be seen in Fig.~\ref{circle_ds}, which is very close to the boundary of the standard trained NN architecture.

The case of the "moons" dataset is a bit different. While the smooth functions are easy to approximate with the nonlinear XY blocks, it is quite complicated to reproduce "sharp" patterns with the high value of the function derivative. For this purpose, we adjust the NN coefficients to achieve good decision boundaries. The difference between the NN and its approximation and consequent results is shown in Fig.~\ref{moon_ds}. Adjusted parameters of NN are given in the \textit{Supplementary Information} section. Figures \ref{circle_ds} and \ref{moon_ds} show the XY blocks of the spin architectures. The presented methodology allows us to upscale the XY blocks with even more complex ML tasks.

\section{Transfering Deep Learning architecture}
\label{Transfering Deep Learning architecture}

Deep NNs are surprisingly efficient at solving practical tasks \cite{sun2017revisiting,lecun2015deep,goodfellow2016deep}. The widely accepted opinion is that the key to this efficiency lies in their depth \cite{kawaguchi2019effect,raghu2017expressive,eldan2016power}. We can transfer the architecture's depth into our XY NN model without any significant loss of accuracy. 

So far, we showed how to transfer the predefined architecture into the XY model. This section discusses the transition of more complex deep architectures, which are considered conventional across different ML fields. To extend the method, we choose two conventional image recognition task models (the architecture details are given in the \textit{Supplementary Information} section).

The focus of the architecture adjustment will be on operations, which were not previously discussed. The list of such operations are $\operatorname{Conv2d}$, $\operatorname{ReLu}$ activation function, $\operatorname{Maxpool2d}$ and $\operatorname{SoftMax}$ \cite{goodfellow2016deep,glorot2011deep,nair2010rectified}. The $\operatorname{Conv2d}$ is a simple convolution operation and does not present a significant difficulty, since it factorizes into the operations previously discussed. $\operatorname{ReLu}$ activation function can be replaced with its analog $\operatorname{LeakyReLu}$ \cite{maas2013rectifier}. Using its similarity with the analytical expression $1.5(1+\tanh{0.8(\theta_{in} - 1.5)})$ allows one to obtain the following set of transformations through $z = F(F((\theta_{in}|-1),(-1.5|-1))|-1),(\pi|1.5))$. Performing the summation of the three terms $1.5$, $F((\theta_{in}|-1),(\pi|-0.9))$ and $-0.5\theta_{in}$ gives us a good approximation for $\operatorname{LeakyReLu}$.

$\operatorname{Maxpool2d}$ relies on  $\operatorname{max}(x,y)$ realisation. Therefore, it is convenient to use two similar architectures of spin value transmission, which are symmetrical with respect to variables $x$ and $y$. The first one consists of $x-y$ operation, $\operatorname{ReLu}$ approximation ( which is essentially $\operatorname{max}(0,x-y)$ function), and summation with the $y$ variable, while the second architecture  interchanges $x$ and $y$ and consists of $y-x$, $\operatorname{ReLu}$ and $+x$ operations. Summing the results of each architecture will give us the required value of $\operatorname{max}(x,y)$.
 
The $\operatorname{SoftMax}$ $e^{z_i}/\sum^K_{j=1} e^{z_j}$ with the input variables $z_i$ can be approximated using several assumptions. Diminishing the exponential embeddings, the problem breaks down into approximating the $\frac{x}{x+y}$ function for $x>0$ and $y>0$. Approximating $\frac{1}{1+0.1y/x} \approx 0.9\tanh(4y/x)$, one can use the block $F((\theta_{in}|-1),(\pi|-0.9))$
to represent $\tanh$ function, while the scaling factor $0.1$ can be controlled by an alternative architecture that depends on  $y$.

\begin{figure} 
\centering
\begin{tabular}{cccc}
\includegraphics[width=0.45\textwidth]{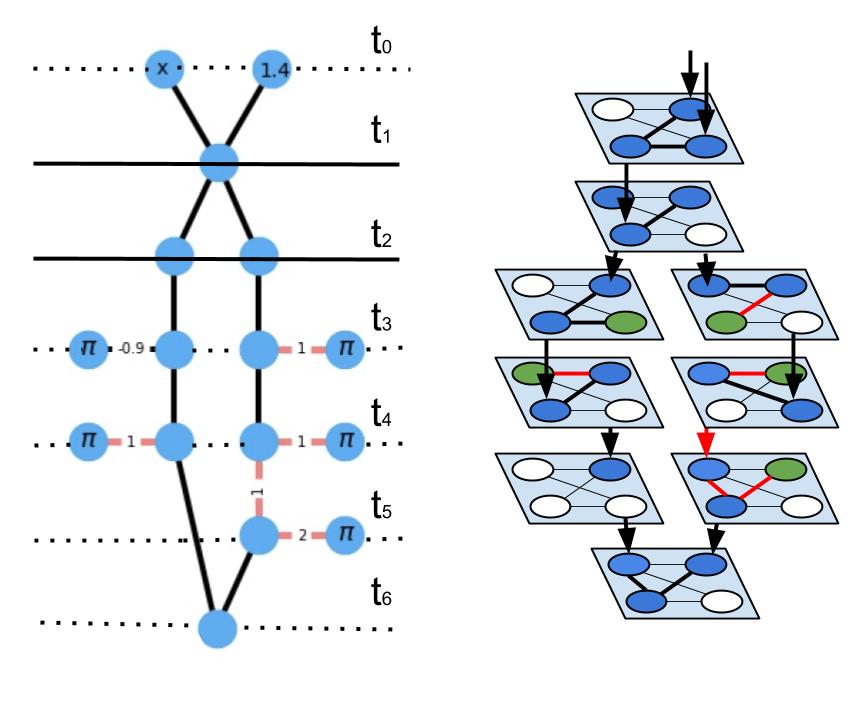} \\
\\[4pt]
\end{tabular}
\caption{The picture's left side demonstrates the graph representation of the $F_t(0.5(x+1.4))$ operation defined in the text. The final spin values are established through $6$ time units, and the measure equals the local characteristic equilibration time of the XY spin cluster. The right side shows the alternative architecture, which performs each operation at the same cluster while saving the local output spin value and transferring it the next time through the self-locking mechanism. The colour correspondence is the same as in XY graphs: blue nodes are the working spins, and green are reference/control spins with $\pi$-values.}
\label{spatial}
\end{figure}

\subsection{Exciton-polariton setting}
\label{Exciton-polariton setting}

In this subsection, we discuss implementing the proposed technique using a system of  exciton-polariton condensates. As we discussed in the Introduction, it is possible to reproduce XY Hamiltonian in the context of exciton-polaritons \cite{berloff2017realizing,lagoudakis2017polariton}, which are hybrid light-matter quasiparticles that are formed in the strong coupling regime in semiconductor microcavities \cite{weisbuch1992observation}.  Each condensate is produced by the pumping sources' spatial modulation and can be treated with the reduced parameters for each unit representing the density and phase degrees of freedom, serving as the analogue variables for the minimization problem.
The exciton-polariton condensate is a gain-dissipative, non-equilibrium system due to the finite quasiparticle lifetimes. Polaritons decay emitting photons. Such emission carries all necessary information of the corresponding system state and can serve as the readout mechanism. Redirecting photons from one condensate to another using an SLM allows to couple the condensates in a lattice  directionally \cite{kalinin2020polaritonic}. 

The system of condensates maximizes the total occupation of condensates by arranging their relative phases so as to minimise  the XY Hamiltonian \cite{berloff2017realizing}. The exciton-polariton platform allows one to manipulate several parameters, such as coupling sterngths between the  condensates to tune them to the particular mathematical operation, or to fix the phase of the condensate  (through the combination of resonant and non-resonant pumping, see \cite{ohadi2016tunable}), and thus creating reference/control spins. Each input spin of the whole system can be controlled via two fixed couplings with the two reference/control spins of different values, see, for example, one of the blocks from Fig.~\ref{basic_plots}.
Fixing the coupling coefficients between other spatially located elements is required to perform the necessary operation and establish the XY network. It can be further upscaled to approximate a particular ML architecture, with the final output spin being the readout target.

One additional note is that the same system can be exploited differently by introducing the spin self-locking mechanism. It consists of saving the spin value in the system without coupling connections with the external elements. This mechanism can be achieved by coupling the local output with another element(s) with high negative coupling strength and consequent decoupling it from the previous units. The self-locking allows one to save the local output and use it for the consequent operations, without significant overhead on elements coming from the previously established operations. We demonstrate the difference in Fig.~\ref{spatial}. The initial universal methodology requires all elements, which the XY graph contains. The presented alternative, requiring a self-locking mechanism, operates with fewer spins by performing each action at the same cluster. Hence, it is volume efficient, which is noticeable in the scaling of elements per operation. Fig.~\ref{spatial} shows the same operation with $16$ spins (without external nodes) and $8$ total spins with the self-locking mechanism. 

\section{Conclusions and future directions}
\label{Conclusions and future directions}

In this paper, we introduced the robust and transparent approach for approximating the standard feedforward NN architectures utilizing the set of nonlinear functions arising from the classical XY spin Hamiltonian behaviour and discussed the possible extensions to other architectures.
The number of additional spins required per operation scales linearly. The best-case scenario has two spin elements per multiplication and nonlinear operation (not taking into account the multiplication by a negative factor), making the general framework rather practical. Some operation approximations used in this work allow additional improvements to reduce the cumulative error between the initial architecture and its nonlinear approximation (see \textit{Supplementary information}).

The entire spectrum of the benefits dramatically depends on a particular type of optical or condensed matter platform. The presented approach has universal applicability, and at the same time, has a certain degree of flexibility. It preserves basic blocks' simplicity, and overall structure works with the intermediate complexity architectures capable of solving toy model data scientific tasks. The upscaling to reproduce DL architectures was discussed.

Our work's side product is a correspondence that allows us to adjust every hardware aiming at finding the ground state of the XY model into the ML task. It becomes possible to build DL hardware from scratch and readjust the existing special-purpose hardware to solve the XY model's minimisation task.

Finally, we would like to  mention the alternative of using a hybrid architecture. Instead of transferring the operations used in the conventional NN model, we can introduce the nonlinear blocks coming from the presented XY model into the working functional given by a particular ML library. For example, we can change the activation function into the one that comes from the system operation and therefore is easily reproducible by that system.
This would result in the  architecture and the transfer process that do not require additional adjustments from the hardware perspective.

The question about the implementation of the backpropagation mechanism, i.e. computing of the gradient of the NN weights in a supervised manner, usually with the gradient descent method, is still under consideration because we limited the scope of the work with the transfer of the predefined, pretrained architecture.

Several possible extensions of our work are possible such as  
extending to k-local Hamiltonians or to other models
with additional degrees of freedom and controls, 
simplifying different mathematical operations and  approximations of the basic functions using many-body clusters in a particular model, increasing the presented approach's functionality (for instance,  adding the backpropagation mechanism).

\section*{Supplementary information. } 

Here we present the parameters of the NNs used in this work and  details of the training and approximations for the XY graphs. We also present two DL architectures mentioned in the main text emphasising their nonlinear operations. Finally, we  discuss the calculations and estimations of the approximation quality.

For training simple classical $(2,3,2)$ feedforward NN architectures on "moons" and "circle" datasets we used the Pytorch library \cite{paszke2017automatic} and Adam optimizer \cite{kingma2014adam}  with batch size $32$ and learning rate $0.01$ value. The expected learning procedure passed without the problems on datasets consisting of $200$ points generated with the small noise of magnitude $0.1$. The final performance gives perfect expected accuracy in both cases.

For the "circles" dataset, the NN parameters are presented in Table \ref{nn_circle_toy}.

\definecolor{mygray}{gray}{0.8}
\begin{table}[!htb]
    \begin{subtable}
      \centering
        \rowcolors[]{1}{}{lightgray}
        \begin{tabular}{lll}
            $w_{11}$ 	& $w_{12}$ & $b_{1}$  \\
            $-1.3465$ 	& $-2.4191$ & $1.1582$ \\
            $-3.5880$ 	& $-0.1474$ & $-1.5228$  \\
            $-1.3565$ 	& $2.6776$ & $1.5239$ \\
        \end{tabular}
    \end{subtable}%
    \begin{subtable}
      \centering
        \rowcolors[]{1}{}{lightgray}
        \begin{tabular}{llll}
            $w_{21}$ 	& $w_{22}$ & $w_{23}$ & $b_{2}$ \\
            $-17.8809$ 	& $16.4797$ & $-18.2794$ & $23.6661$ \\
            $17.3684$ 	& $-17.0227$ & $18.4634$ & $-23.3256$  \\
        \end{tabular}
    \end{subtable}
    \caption{NN (2,3,2) parameters used for the toy dataset 'circles.'}
    \label{nn_circle_toy}
\end{table}

The first row of mathematical approximations gives us similar coefficients. We can rewrite in the same manner the NN parameters with minor adjustments for demonstrative purposes (see the main text for the detailed analysis of possible assumptions and the improvements such as  the multiplication by the scaling coefficients).

\begin{table}[!htb]
    \begin{subtable}
      \centering
        \rowcolors{1}{white}{mygray}
        \begin{tabular}{lll}
            $w_{11}$ 	& $w_{12}$ & $b_{1}$  \\
            $-0.5$ 	& $-0.75$ & $0.35$ \\
            $1.0$ 	& $0.0$ & $0.45$  \\
            $-0.5$ 	& $1.0$ & $0.6$ \\
        \end{tabular}
    \end{subtable}%
    \begin{subtable}
      \centering
        \rowcolors{1}{white}{mygray}
        \begin{tabular}{llll}
            $w_{21}$ 	& $w_{22}$ & $w_{23}$ & $b_{2}$ \\
            $1.0$ 	& $1.0$ & $1.0$ & $-0.31$ \\
        \end{tabular}
    \end{subtable}
    \caption{XY NN (2,3,1) parameters used to approximate the standard NN for the toy dataset 'circles'.}
    \label{nn_xy_circle_toy}
\end{table}

The presented approximation architecture given in Table \ref{nn_xy_circle_toy} was adjusted for better representation of  one final feature, which is general enough to mark the decision boundaries for this particular task, while getting rid of the unnecessary parameters.
Another approximation stage leads us to the final architecture, which is shown  in Fig.~\ref{circle_ds}. Let $f_{i}$ denote the $i$-th feature and $R(x) = F((F_1(x)|-1),(F_3(F_2(x))|-1))$, where $F_1(x) = F((x|-1),(\pi|-0.9)), F_2(x) = F((x|-1),(\pi|1)), F_3(x) = F((x|1),(\pi|2))$ represents the approximation of the activation function with the reduced accuracy, then:\newline
$x_{11} = F((F((x_1|-1),(\pi|0.99))|1),(\pi|2))$\newline
$x_{12} = F((F((x_{11}|-1),(\pi|0.99))|1),(\pi|2))$\newline
$y_{11} = F((F((y_1|-1),(\pi|0.99))|1),(\pi|2))$\newline
$y_{12} = F((F((y_{11}|-1),(\pi|-0.08))|-1),(\pi|-0.08))$\newline
$f_1 = F((x_{12}|-1),(y_{12}|-1),(b_1=0.35|-1))$\newline
$f_{11} = R(f_1);$\newline
$x_{21} = F((F((x_2|-1),(\pi|0.99))|1),(\pi|2))$\newline
$f_2 = F((x_{21}|-1),(y_{2}|-1),(b_2=0.6|-1))$\newline
$f_{22} = R(f_2);$\newline
$f_3 = F((x_{3}|-1),(y_{3}|0),(b_3=0.45|-1))$\newline
$f_{33} = R(f_3);$\newline
$G_0 = F((f_{11}|-1),(f_{22}|-1),(f_{33}|-1))$\newline
$G = F((G_{0}|-1),(g=-0.1033|-1))$\newline

This structure is  represented in  Fig.~\ref{circle_ds}.

The NN parameters for the "moons" dataset  are  presented in \ref{nn_moons_toy} Table.

\definecolor{mygray}{gray}{0.8}
\begin{table}[!htb]
    \begin{subtable}
      \centering
        \rowcolors{1}{white}{mygray}
        \begin{tabular}{lll}
            $w_{11}$ 	& $w_{12}$ & $b_{1}$  \\
            $6.2888$ 	& $-3.2930$ & $-3.0992$ \\
            $-3.5880$ 	& $-4.2940$ & $5.9965$  \\
            $-6.1958$ 	& $-2.7684$ & $0.6882$ \\
        \end{tabular}
    \end{subtable}%
    \begin{subtable}
      \centering
        \rowcolors{1}{white}{mygray}
        \begin{tabular}{llll}
            $w_{21}$ 	& $w_{22}$ & $w_{23}$ & $b_{2}$ \\
            $-6.1143$ 	& $-6.8860$ & $-6.8151$ & $-0.0621$ \\
            $6.6825$ 	& $6.2848$ & $6.8381$ & $-0.0325$  \\
        \end{tabular}
    \end{subtable}
    \caption{NN (2,3,2) parameters used for the toy dataset 'moons.'}
    \label{nn_moons_toy}
\end{table}

First row of the  approximations parameters is given in Table \ref{nn_moons_toy} .

\begin{table}[!htb]
    \begin{subtable}
      \centering
        \rowcolors{1}{white}{mygray}
        \begin{tabular}{lll}
            $w_{11}$ 	& $w_{12}$ & $b_{1}$  \\
            $1.0$ 	& $-0.125$ & $-0.9$ \\
            $1.0$ 	& $-0.125$ & $0.9$  \\
            $-0.5$ 	& $-0.125$ & $0$ \\
        \end{tabular}
    \end{subtable}%
    \begin{subtable}
      \centering
        \rowcolors{1}{white}{mygray}
        \begin{tabular}{llll}
            $w_{21}$ 	& $w_{22}$ & $w_{23}$ & $b_{2}$ \\
            $1.0$ 	& $1.0$ & $1.0$ & $0.065$ \\
        \end{tabular}
    \end{subtable}
    \caption{XY NN (2,3,1) parameters used to approximate the standard NN for the toy dataset 'moons.'}
    \label{nn_xy_moons_toy}
\end{table}

The presented  architecture was adjusted for better representation of  one final feature. For the case of "moons," the additional adjustment has been added since the presented XY architecture has lower expressivity for the case of sharp boundaries.
Another approximation stage leads us to the final architecture, which can be found in Fig.~\ref{moon_ds}:\newline
$y_{11} = F((F((y_1|-1),(\pi|0.99))|-1),(\pi|0.99))$\newline
$y_{12} = F((F((y_{11}|-1),(\pi|0.99))|1),(\pi|2))$\newline
$f_1 = F((x_{1}|-1),(y_{12}|-1),(b_1=-0.95|-1))$\newline
$f_{11} = R(f_1);$\newline
$y_{21} = F((F((y_2|-1),(\pi|0.99))|-1),(\pi|0.99))$\newline
$y_{22} = F((F((y_{21}|-1),(\pi|0.99))|1),(\pi|2))$\newline
$f_2 = F((x_{1}|-1),(y_{22}|-1),(b_1=0.9|-1))$\newline
$f_{22} = R(f_2);$\newline
$x_{31} = F((F((x_{3}|-1),(\pi|0.99))|1),(\pi|2))$\newline
$y_{31} = F((F((y_{3}|-1),(\pi|0.99))|-1),(\pi|0.99))$\newline
$y_{32} = F((F((y_{31}|-1),(\pi|0.99))|1),(\pi|2))$\newline
$f_3 = F((x_{31}|-1),(y_{32}|-1),(b_1=0|-1))$\newline
$f_{33} = R(f_3);$\newline
$G_0 = F((f_{11}|-1),(f_{22}|-1),(f_{33}|-1))$\newline
$G = F((G_{0}|-1),(g=0.0216|-1))$\newline

The presented structure follows  the  graph structure given in Fig.~\ref{moon_ds}.

The presented DL architectures are defined with the Pytorch library's help in the following Table \ref{dl_parameters}.

\begin{table}[!htb]
    \begin{subtable}
      \centering
        \rowcolors{1}{white}{mygray}
        \begin{tabular}{ll}
            NN layer \\
            5 $\times$ 5 Conv2d(3,6) \\
            2 $\times$ 2 MaxPool2d \\
            5 $\times$ 5 Conv2d(6,16) \\
            Linear (400,120) \\
            Linear (120,84) \\
            Linear (84,10) \\
        \end{tabular}
    \end{subtable}%
    \begin{subtable}
      \centering
        \rowcolors{1}{white}{mygray}
        \begin{tabular}{llll}
            NN layer \\
            5 $\times$ 5 Conv2d(1,10) \\
            2 $\times$ 2 MaxPool2d \\
            ReLu \\
            Dropout(0.5) \\
            5 $\times$ 5 Conv2d(10,20) \\
            2 $\times$ 2 MaxPool2d \\
            ReLu \\
            Flatten \\
            Linear (320,50) \\
            ReLu \\
            Linear (50,10) \\
            Softmax \\
        \end{tabular}
    \end{subtable}
    \caption{Examples of the simple DL architectures used to represent nonlinear/unique functions. (a) NN for simple 10 classes digit recognition. (b) NN for CIFAR10 dataset classification.}
    \label{dl_parameters}
\end{table}

In Table \ref{dl_parameters}, we present the details of two DL architectures that were discussed in the main text emphasizing their nonlinear operations.

Finally, we show that the initial task of approximating a particular set of mathematical operations by the parametrized family of nonlinear functions can be done more rigorously with a potential for the accumulated error estimation through the layers of NN. 

The discrepancy between the target function and its approximation can be estimated with the $L^1([-\pi /2,\pi /2])$ norm on the working domain:
\begin{eqnarray}
L^1&=& \int_{-\pi /2}^{\pi /2}|-\operatorname{sign}B(x,\{J_i\}) \nonumber\\&& \biggl(\frac{\pi}{2} +  \arcsin\frac{A(x,\{J_i\})}{\sqrt{A(x,\{J_i\})^2 + B(x,\{J_i\})^2}}\biggr)\nonumber\\ &-&f(x)_{\rm target}|dx.
\label{L1_norm_discrepancy}		
\end{eqnarray}
Starting with the multiplication operation, one can calculate Eq.~(\ref{L1_norm_discrepancy}) with  $f(x,k)_{target} = kx$ and obtain the expression (depending on the $(J,k)$ parameters) for one block of spins. Further minimisation of Eq.~(\ref{L1_norm_discrepancy}) leads to the expression for  $J(k).$  Evaluating Eq.(\ref{L1_norm_discrepancy}) analytically can be done in simpler way by replacing the expression involving $\operatorname{arcsin}$ function the one with $\operatorname{arcctg}$, so that initial integral (in terms of the argument of the complex parameter $C = \sum^{N-1}_{i} J_{i} e^{i \theta_{i}}$) contains the following  expression for one input and one control/reference spin:
\begin{equation}
\scalebox{0.9}{
$I=\int_{-\pi /2}^{\pi /2} \operatorname{arcctg}\frac{B((x_|-1),(\pi|J))}{A((x_|-1),(\pi|J))}dx = 
\int_{-\pi /2}^{\pi /2} \operatorname{arcctg}\frac{\sin(x)}{J+\cos(x)}dx.$}
\label{arcctg}		
\end{equation}
We evaluate this to
\begin{eqnarray}
I=x\operatorname{arcctg}\frac{\sin(x)}{J+\cos(x)} +\frac{1}{4}(x^2+2i\operatorname{sign}(J^2-1)\times\nonumber \\
\times(i[\operatorname{Li}_2(\frac{D(1-E)}{1+E})+\operatorname{Li}_2(\frac{D^{*}(1-E)}{1+E})] +\nonumber \\ +2x\operatorname{arctanh}(E^{-1})-G\operatorname{arctanh}(E) +\nonumber \\
+[G-2i\operatorname{arctanh}(E)]\log\frac{2J(1+E)}{I_1(tg(x/2)-i)} +\nonumber \\
+[G+2i\operatorname{arctanh}(E)]\log\frac{2J(1+E)}{I_2(tg(x/2)+i)} +\nonumber \\
+\log He^{-ix/2}(2i\operatorname{arctanh}(E)-2i\operatorname{arctanh}(E^{-1})+G) +\nonumber \\
+\log He^{ix/2}(2i\operatorname{arctanh}(E^{-1})-2i\operatorname{arctanh}(E)+G)))\bigg|_{-\pi /2}^{\pi /2}, \nonumber \\
\label{analytical_expression}		
\end{eqnarray}
with variables $D(J) = (J^2+1+|J^2-1|)/2J$, $E(x,J)=\frac{i|J^2-1|}{(J+1)^2}\operatorname{tg}x/2$, $C(J) = \operatorname{arccos}(-\frac{J^2+1}{2J})$, $H(J) = \frac{i|J^2-1|}{2\sqrt{J}\sqrt{J^2+2J\cos{x}+1}}$,$I_1(J)=2i(J-1), \operatorname{if} J^2>1; 2iJ(1-J),\operatorname{otherwise}$, $I_2(J)=2iJ(J-1), \operatorname{if} J^2>1; 2J(J-1),\operatorname{otherwise}$ and $\operatorname{arcctg},\operatorname{arctanh},\operatorname{arccos}$ denoting the inverse for tangent, hyperbolic tangent, cosine functions respectively with $\operatorname{Li}_s(x) = \sum_{k=1}^{\infty}x^k/k^s$ being the polylogarithm function and $*$ denoting the complex conjugate operation. One can simplify the given formula by the contraction of the complex pair terms:
\begin{align}
I=x\operatorname{arcctg}\frac{\sin(x)}{J+\cos(x)} +\frac{1}{4}(x^2+2i\operatorname{sign}(J^2-1)\times \nonumber \\ \times (i[\sum_{k=1}^{\infty}\frac{2\cos(k\phi)}{k^2}] +2G\log H \nonumber +(2x-G)\operatorname{arctanh}(E) +\nonumber \\
+ G\log\frac{J(1+E)^2}{E^2(J+1)^2-(J-1)^2} -2i \operatorname{sign}(J^2-1) \times \nonumber \\
\times \operatorname{arctanh}(E) \log \frac{J(E(J+1)-(1-J))}{E(J+1)-(J-1)}))\bigg|_{-\pi /2}^{\pi /2},
\label{simplified_analytical_expression}		
\end{align}
where we added a new variable $\phi = \operatorname{Arg(D(1-E)/(1+E)}$. 

\begin{figure} 
\centering
\includegraphics[width=0.48\textwidth]{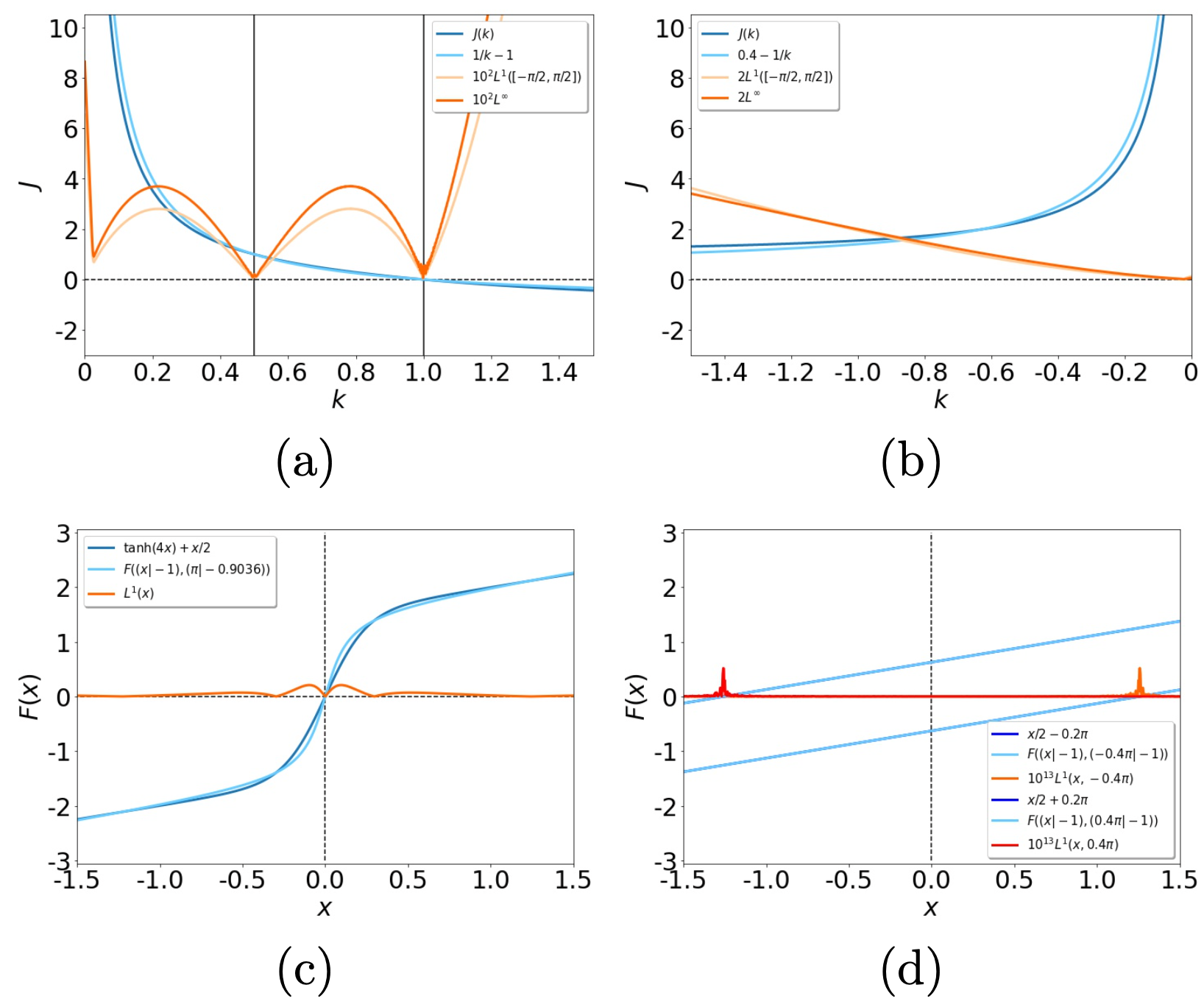}
\caption{Top: (a) Function $J(k)$ that minimizes Eq.~(\ref{L1_norm_discrepancy}) with $F((x|-1),(\pi|J))$,  $f(x)_{target} = kx$ and a positive factor $k$ (blue line),  its approximation (white blue line) and the fitted formula $\approx 1/k - 1$. The supporting plots depict  $10^{2} L^{1}$ (light orange line) and $10^{2} L^{\infty}$ (orange) for each value of $k$. Vertical black lines denote
the points with the minimal accumulated error.
(b) Function $J(k)$  that minimises Eq.~(\ref{L1_norm_discrepancy}) with $F((x|1),(\pi|J))$,  $f(x)_{target} = kx$ and a negative factor $k$ (blue line),  its approximation (white blue line) and the fitted formula $\approx 0.4 - 1/k$. The supporting plots depict $2 L^{1}$ (light orange line) and $2 L^{\infty}$ (orange) for each value of $k$.
Bottom: (c) Function $\tanh{4x}+x/2$ (blue line) and its approximation with the function $F((x|-1),(\pi|-0.9036))$ (light blue line) with the optimised parameter $J$. The supporting orange plot represents the error at each point on $x$-axis.
(d) The half-sums for $x$ and two variables $y = -0.2\pi,0.2\pi$ (blue lines), which coincide with their approximations $F((x|-1),(y|-1))$ (white blue) giving insignificant approximating error $10^{13} L^{1}$ (red and orange lines).
}
\label{error_plots}
\end{figure}

To shorten the description of the  dependence of the coupling strength $J$ on the multiplication factor $k$ and avoid overcomplicated analytical expressions, we present the plot of its approximation, which alternatively can be calculated numerically and can be approximated  by an expression $1/k - 1$ with a good accuracy. Additionally, we calculated  $L^1([-\pi /2,\pi /2])$  according to Eq.~ (\ref{L1_norm_discrepancy}) for each value of $k$. Another good measure of the approximation quality is $L^{\infty}$, which is the maximal discrepancy between the functions $F((x|-1),(\pi|J))$ and $f(x)_{target}$, which has a similar behaviour
as the original norm. Fig.~(\ref{error_plots}) (a) shows all the plots corresponding to the multiplication by a positive factor $k>0$ with the special points of the minimal error at $k=1,0.5$ and $k=0$. These graphs explain  why the lesser factors are more reliable for the multiplication, and why multiplying by a larger factors without  factorization leads to worse performance.

The same task of multiplication, but by a negative factor $k<0$ is illustrated in Fig.~(\ref{error_plots})(b). The $J(k)$ function can be approximated with a reasonable accuracy by an expression $0.4 - 1/k$. Since the general error has a much higher factor $\approx 10^{2}$ for the negative values, one has to accompany this block with an additional linear embeddings to achieve a good approximation. The nonlinear function $3/2 \tanh{4x}+x/2$ and its approximation with the optimised function $F((x|-1),(\pi|-0.9036))$ is depicted in the Fig.~(\ref{error_plots})(c). The lowest accuracy is observed near the origin.

The final example of the half-sum approximation is illustrated in Fig.~(\ref{error_plots})(d). The surprisingly good agreement (with an error of $10^{-13}$ of the magnitude) between the initial function and its XY representation  $F((x|-1),(y|-1))$ can be explained with the help of the Taylor expansion of  Eq.(\ref{xy_hamiltonian_analytics}) near zeros. It gives the linear coefficient of  $1/2$  accurate to the fourth order of approximation. With all the presented information, one can estimate the maximal discrepancy between an arbitrary NN and its transferred XY analog, which will be a good measure of the approximation quality and final adequacy of the transfer.

\bibliography{Bibliography}{}
\bibliographystyle{ieeetr}

\end{document}